# Integrating Generative AI into LMS: Reshaping Learning and Instructional Design

Xinran Zhu[1], Liam Magee[1], Peg Mischler[2]
[1]University of Illinois Urbana Champaign
[2]CheckIT Learning

Education in the era of generative AI will undergo dramatic change. Education prepares students for future work, and already, AI appears to be preparing a very different world of employment. In the field of software, for example, tools such as Anthropic's Claude Code, OpenAI's Code, and Google's Gemini CLI modify how code is written, debugged, and deployed. Does this make traditional computer science obsolete? No, but it does suggest new ways in which software will be designed and maintained—and if it is to prepare future workers, teaching needs to adjust accordingly. In particular, the design and development of AI tools should be grounded in an understanding of social and ethical challenges, as well as in theories of learning and cognition. At the same time, preparing students for this future requires more than technical know-how; it calls for the ability to engage critically and meaningfully with such intelligent systems. To equip students for the realities of tomorrow's workforce, the design of AI tools must be grounded not only in technology but also in theories of learning and cognition. Here, the role of the Learning Management System (LMS) becomes vital: as the hub of instruction, it must evolve to embed generative AI in ways that both enhance pedagogy and prepare learners for a new world of work.

Various technologies have long supported education. From the abacus to calculators, personal computers, projectors, and eventually web-based LMSs, technology has always served as a medium for teacher instruction and student cognition, memorization, and practice. Video lecture platforms like Khan Academy have made it possible to learn remotely, while gamified apps like Duolingo provide smooth pathways for students who are deterred from learning through traditional textbook methods. Yet even these examples assume a relatively static profile of the learner. Content is delivered and assessed according to assumptions made—typically by expert teachers—about the appropriate pace and learning tendencies.

Students vary widely in their approaches to learning and in how their learning progresses over time. Some require repeated exposure to material before it "clicks," and once it does, their progress accelerates dramatically—Einstein's famously slow initial progress in science exemplifies this pattern. Some students move quickly through the early stages of mastery, only to stall at a critical point—and, out of frustration, make little further progress. This is often seen when learners reach the challenges of calculus, Shakespearean drama, or the complexities of verb tenses in a new language. In most classrooms, teachers lack the time and resources to adapt to these nonlinear progressions. At the same time, traditional software remains too rigid to adjust to the dynamic realities of how learning unfolds.

Generative AI combines many of the affordances of recent IT innovation with extreme customization. No longer must educational materials be delivered in a one-size-fits-all model. These materials can be delivered at different paces, in variable sequences, and with activities and assessments that are precisely tailored to individual students' zone of proximal development – the optimal point where learning is challenging enough to inspire growth without being discouraging. In this way, AI has the potential to keep learners engaged, motivated, and progressing steadily toward mastery.

However, generative AI tools such as ChatGPT and Claude are delivered in the form of generic chatbots, and these formats are rarely optimized for learning environments. These tools need to be scaffolded within learning systems that can coordinate content across learning levels, classes, and curricula. Systems like CheckIT Learning—a neuroscience-informed learning ecosystem built for teachers, administrators, and students—demonstrate how generative AI can be integrated into a structured and scalable approach to pedagogy. Other emerging systems, such as CyberScholar, integrate generative AI into the actual processes of writing—not by writing essays on the students' behalf, but by reviewing, raising questions, and offering qualified support that assists, rather than diminishes, the experience of learning through composition.

## From Content Delivery to Fostering Higher-Order Thinking

Much of the excitement around AI in education has centered on personalization and automation. Yet AI also holds the potential to foster higher-order thinking skills—such as collaboration, critical thinking, and reflection—that traditional instruction often struggles to cultivate. In today's rapidly evolving world, it is no longer sufficient for students to passively absorb information, especially when AI can easily generate that information. Instead, they must become knowledge creators, capable of collaboratively advancing ideas toward deeper understanding and innovation (Scardamalia & Bereiter, 2006). To cultivate these capacities, students must engage in intentional and reflective inquiry, both individually and collaboratively with others. Such inquiry involves posing meaningful questions, identifying knowledge gaps, building on existing ideas, integrating diverse perspectives, and iteratively refining shared artifacts within a community of learners.

These knowledge-building practices prepare students not only to navigate complexity, but also to shape it—whether in addressing climate change, developing responsible technologies, or tackling unforeseen global challenges. Within this paradigm, AI should not be treated as a shortcut to answers, but as a partner in inquiry: helping learners surface assumptions, explore alternatives, and articulate emerging understandings.

## Toward Meaningful Interaction with AI

Meaningful interaction with AI requires more than tool use. It demands critical, reflective, and socially mediated engagement across the learning processes. As Shibani et al. (2024) argue, learners often interact with AI in shallow, task-driven ways. Their Critical Interaction with AI for Writing (CIAW) framework identifies several dimensions of deep engagement with AI, including thoughtful use during ideation, critical analysis of information, productive dialogue with AI, and reflective evaluation of its role in the learning process. The challenge is to design environments that prompt students to use AI critically and thoughtfully, rather than simply as a tool for automatic answers.

At the heart of meaningful human–AI interaction is the recognition that humans and AI bring distinct yet complementary capabilities to the learning process (Rees, 2022). Humans are sense-makers, intentional agents, and ethical decision-makers. We formulate meaningful questions, pursue evolving goals, and make judgments rooted in context, experience, and personal values. These capacities enable us to interpret ambiguity, weigh competing perspectives, and respond creatively to uncertainty—skills that are essential not only for academic success but also for navigating complex, real-world challenges. In educational settings, such capacities are key to how learners engage with knowledge. They form the foundation of epistemic agency: the ability to decide what is worth knowing, initiate inquiry, and refine understanding through critical reflection and dialogue.

AI is well-suited for tasks that involve recognizing patterns, retrieving information, and generating fluent, well-structured language. Tools like large language models can quickly summarize readings, suggest counterarguments, or propose examples based on statistical associations in their training data. They can stimulate epistemic moves; however, AI lacks intentionality, personal goals, and a grasp of meaning—it does not know what is important or why something matters (at least for now).

The complementary strengths of human and AI can be leveraged to support sustained, reflective, and purposeful inquiry and creative knowledge building. For instance, teachers use AI-generated insights not as fixed answers, but as resources that they interpret through the lens of their pedagogical goals and professional identities, tailoring them to their students' specific needs. Students also play an active role by evaluating whether AI responses align with their intentions and initiating the next steps. This aligns with the concept of hybrid intelligence (Dellermann et al., 2019), which advocates for the purposeful combination of human and machine capabilities to augment human learning and problem-solving. In a mindful human–AI partnership, AI supports what it does best: summarizing, organizing, suggesting, and modeling ideas. It acts as a cognitive scaffold, surfacing gaps, offering alternatives, or prompting new directions. Humans, in turn, guide the inquiry with purpose and discernment—evaluating not just the content of the AI's suggestions, but also the reasoning, relevance, and meaning behind them.

## Case Study on Integrating AI in LMS: CheckIT Learning

In this section, we describe how CheckIT Learning, an edtech startup, is working to deliver meaningful AI interactions through its Learning Management System. As founders Sava Opacic and Srdan Perovic explain, CheckIT moves beyond viewing AI as a mere tool. Grounded in decades of research from neuroscience, cognitive science, and learning sciences, the platform is designed to advance responsible AI in education. Its AI mentor, Cleo, adapts to different rates of learning by calibrating support as students encounter new material, build competency, and ultimately achieve mastery. For example, Cleo provides students with personalized study strategies and timely feedback that connect effort to progress, while supporting teachers during lesson planning by embedding evidence-based practices directly into units, pacing, and assessments. Teachers receive suggestions on where to insert retrieval practice, attention re-engagement, or formative checks, ensuring lessons align with how the brain learns most effectively. In this way, students experience greater confidence and motivation, and teachers gain clarity and time to focus on the art of teaching rather than routine design tasks.

One example of how research translates into practice is in the design of lessons that sustain attention. Classic and contemporary work (e.g., James, 1890; Broadbent, 1958; Posner & Petersen, 1990) shows that attention is limited and dynamic. Research on classroom learning (e.g., Souza, 2022) further suggests that student focus fluctuates—often higher at openings or after transitions, but lower during extended passive listening. Cleo helps teachers respond to these natural rhythms by recommending shorter, varied activities that require students to retrieve, test, and reflect on earlier content (Karpicke & Roediger, 2008; Sousa, 2022). These activities are ideally delivered in multimodal formats that reduce cognitive load (Johnson & Mayer, 2009).

Cleo's role extends beyond moment-to-moment engagement. The platform provides teachers with actionable insights during lesson planning and delivery, highlighting where to embed retrieval practice, attention re-engagement, or formative checks to ensure lessons align with how the brain learns best. For students, Cleo connects daily learning activities to their broader goals through the Vision Board, helping them see progress in relation to passions, values, and career aspirations. Its analytics surface patterns of motivation, persistence, and progress over time, giving both students and teachers a clearer picture of growth. In this way, students strengthen metacognition—learning how to reflect on effort, strategies, and outcomes—while teachers gain confidence that their instructional design and feedback are tied to research-based practices. This approach reflects research on the importance of purpose (Damon, 2008; Yeager et al., 2014) and self-regulation (Duckworth & Gross, 2014; Blair & Raver, 2015), ensuring that classroom experiences are not only efficient but also meaningful and future-oriented.

However, technology implementations that support these comparatively new findings in neuroscience and learning science – such as supporting attention re-engagement, embedding multimedia learning principles, cultivating a sense of purpose, and strengthening student self-regulation—remain scarce. Most learning management systems provide little more than general scaffolds for content, requiring instead that teachers design and deliver that content. AI-augmented platforms like CheckIT Learning offer both flexibility and support: teachers can create and teach lessons according to their own philosophy, student needs, learning levels, and disciplinary demands, while also leveraging automation to provide personalized, research-aligned learning experiences.

A key site of human-AI collaboration is CheckIT Learning's lesson planning feature. Teachers begin by selecting their preferred instructional approach—such as direct instruction, collaborative learning, inquiry, project-based work, lecture, or lab-based exploration—and Cleo responds with drafted lesson structures that include objectives, suggested activities, accommodations, and checks for understanding. The process is intentionally iterative and collaborative: teachers refine and adapt Cleo's recommendations based on their professional judgment, the classroom context, and the students' needs. This dynamic exchange exemplifies how human and AI agents can co-create instructional design, each contributing distinct strengths. While Cleo provides efficiency, alignment with neuroscience and learning science research, and access to large-scale patterns, the teacher ensures contextual insight, creativity, and pedagogical nuance remain at the center of the lesson.

Cleo's role in student content creation is intentionally collaborative. Once teachers finalize their lesson plans, Cleo generates student-facing content designed with neuroscience in mind—framing introductions to activate curiosity, lowering anxiety through encouraging tone, sequencing objectives to guide attention, and presenting core ideas in formats that support dual coding and working memory. Teachers refine this content, adapting examples, pacing, and tone to their classroom context and students' lived experiences. In this way, AI offers efficiency and alignment with research, while teachers ensure relevance, authenticity, and cultural connection.

This workflow goes beyond automation by engaging teachers in intellectually rich design processes that promote students' disciplinary understanding and critical reflection. At the same time, it builds teacher capacity by modeling pedagogically grounded strategies and increasing their AI literacy. Through repeated, reflective use, teachers not only co-design more meaningful learning experiences but also become better equipped to engage with AI tools in future educational contexts critically.

## The Need for Edtech Partnerships in an AI-Powered World

Today, the AI landscape for learning remains very immature. The relentless release of new foundation models is expected to continue into 2025. The recent release (July 2025) of ChatGPT's *Study Mode* also demonstrates that the creators of these models are shifting their focus to education. A sign, too, that the world of AI-powered apps for learning is now upon us.

These models mobilize rounds of new technological innovation. Still, at the same time, for teachers and educators, this innovation can also lead to superficial "AI-fication"—the inclusion of AI everywhere, without a clear purpose or value. Just as with other innovations – such as smartphones, games, and social media – AI technology can also be highly corrosive of attention, retention, and critical thinking.

To date, research on how machine learning can be applied meaningfully to human learning, in its immense diversity, is suggestive but unclear. Of special concern is the thin line that separates so-called "cognitive debt" (Kosmyna et al. 2025) – getting AI to do a student's work – from genuine cognitive development (Floridi, Morley, & Novelli, 2025). What exactly happens when a learner interacts with AI? Do they simply enter a query and copy and paste the response? Do they instead read and internalize the result? Do follow-up queries indicate learning? How can AI interactions prompt metacognitive and reflective activity? Is this answer correct? Why is the machine telling me this? How do I reconcile the answer with everything else I know? And from a more philosophical point of view, we also need to ask: Does AI mean we need to revise "classical" psychological and neurological theories of childhood development? Such questions require in-depth research and practical application.

In the current AI gold rush, it often appears as though technology providers lack capacity – or interest – to undertake this kind of "deep learning". Meanwhile, universities lack the capital and the talent to build experimental learning platforms. Corporate-academic partnerships – long integral to genuine innovation – are needed to understand the more profound questions posed by AI's arrival into education. In parallel, education research has long demonstrated this potential through research–practice partnerships (RPPs), which show how long-term, mutualistic collaborations between researchers and practitioners can tackle complex problems of practice. By emphasizing shared authority, sustained engagement, and the co-design of solutions, this type of partnership or collaboration fosters conditions that enable both rigorous inquiry and practical impact (Coburn & Penuel, 2016).

These opportunities and challenges call for the mindful integration of AI in education, grounded in theories of learning, teaching, and cognition. This work should be driven by collaboration among researchers, educators, technologists, and policymakers, with the aim of advancing learner agency, equity, and innovation in a rapidly evolving society. Companies like CheckIT Learning are combining AI with neuroscience research to create new symbiotic links between machine and human learning. Meanwhile, foundational research in neuroscience, learning sciences, sociology, and science and technology studies is exploring how individuals, communities, and industries can leverage the possibilities of AI – while also assessing its risks. At the intersection of these developments lies the opportunity to accelerate and diversify human learning at new levels.

## Acknowledgements

While the views expressed in this white paper are our own, we acknowledge the significant time and effort from CheckIT Learning in helping to prepare the information and conclusions presented.